\documentclass[a4paper,11pt]{article}
\usepackage{pos}
\usepackage[acronym]{glossaries}
\glsdisablehyper
\usepackage{graphicx}
\usepackage{wrapfig}
\usepackage{multicol}
\usepackage{slashed}
\usepackage{physics}
\usepackage{subcaption}
\usepackage{wrapfig}
\usepackage{dsfont}
\usepackage[colorinlistoftodos,prependcaption,textsize=small]{todonotes}

\definecolor{purple}{rgb}{0.5,0.0,0.5}

\newacronym{rschipt}{rS$\chi$PT}{Rooted Staggered Chiral Perturbation Theory}
\newacronym{schipt}{S$\chi$PT}{Staggered Chiral Perturbation Theory}
\newacronym{chipt}{$\chi$PT}{Chiral Perturbation Theory}
\newacronym{lqcd}{LQCD}{Lattice QCD}
\newacronym{lec}{LEC}{Low-Energy Constant}
\newacronym{bs}{BS}{Bethe-Salpeter}
\newacronym{cm}{CM}{Center of Mass}
\newacronym{lo}{LO}{Leading Order}
\newacronym{nlo}{NLO}{Next-to-Leading Order}

\newcommand{\M}{\mathcal{M}}

\newcommand{\I}{\mathcal{I}}
\renewcommand{\S}{\mathcal{S}}
\renewcommand{\vec}[1]{\textbf{#1}}

\title{Towards a formalism for $\pi\pi$ scattering from staggered lattice QCD}

\author*[a]{A. Dean. M. Valois}
\author[b]{M. Dai}
\author[b]{A. El-Khadra}
\author[a]{E. Gámiz}
\author[c]{S. Lahert}
\author[a]{R. Merino}

\affiliation[a]{CAFPE and Departamento de Física Teórica y del Cosmos, Universidad de Granada, E-18071
Granada, Spain}

\affiliation[b]{Department of Physics, University of Illinois, Urbana, Illinois, 61801, USA}

\affiliation[c]{Department of Physics and Astronomy, University of Utah, Salt Lake City, UT, USA}

\emailAdd{deanvalois@ugr.es}

\abstract{

\vspace*{-2mm}
\textbf{\textsf{Fermilab Lattice and MILC Collaborations}}\\[0.7em]
Scattering processes featuring the strong interactions can be studied using lattice QCD by means of the Lüscher formalism. This approach relies on analyticity and unitarity of the $S$-matrix to relate infinite-volume scattering amplitudes to finite-volume energy levels. However, lattice QCD simulations employing rooted staggered fermions manifest unitarity violation as an $\mathcal{O}(a^2)$ lattice artifact. Moreover, the meson sector of this theory contains multiple non-mass-degenerate pions (due to the so-called taste splitting), which only reduce to the physical pion in the continuum limit. These features restrict the applicability of the Lüscher formalism to staggered lattice data at non-zero lattice spacing. Hence, in this work, we discuss two complementary approaches to deal with the challenges of extracting $\pi\pi$ scattering amplitudes from lattice QCD with staggered quarks: (1) using the corresponding effective theory, Rooted Staggered Chiral Perturbation Theory, to calculate one-loop amplitudes for the first time. These amplitudes can be used to explicitly check the validity of the quantization condition. And (2) generalizing the formalism to incorporate taste-splitting as well as fourth-rooting effects. We focus on the simpler case of $\pi\pi$ scattering in the isospin-2 channel, and discuss prospects for other channels.}

\FullConference{The 42nd International Symposium on Lattice Field Theory (LATTICE2025)\\
2-8 November 2025\\
Tata Institute of Fundamental Research, Mumbai, India\\}

\begin{document}
\maketitle

\section{Introduction}\label{sec:intro}
Scattering amplitudes in various physical processes governed by the strong interactions can be computed from \gls{lqcd} data via the Lüscher formalism. Specifically, the formalism connects discrete energy spectra of particles in Euclidean finite-volume spacetime with the corresponding infinite-volume scattering amplitudes in Minkowski spacetime by means of the so-called quantization condition. Since the seminal work by Martin Lüscher~\cite{Luscher:1986pf} -- which considered the scattering of two identical scalar particles in the \gls{cm} frame -- various extensions of the formalism have been developed to include spin quantum numbers, coupled scattering channels, moving frames, 3-body scattering, etc. Altogether, the Lüscher formalism and its extensions have been remarkably successful in unveiling hadronic properties, including, for instance, those of stable bound states and resonances (see Ref.~\cite{Briceno:2017max} for a broad review, and Ref.~\cite{Green:2026jgv} for an overview of recent progress presented in this conference).

The input energies required by the quantization condition are typically computed within the framework of \gls{lqcd}, where the finite volume plays a key role in the scattering formalism. Besides finite-volume effects, lattice simulations are also inherently affected by discretization effects, which scale with a given power of the lattice spacing $a$. Nevertheless, these effects are not taken into account in the standard Lüscher formalism. Recently, this issue was addressed in Ref.~\cite{Hansen:2024cai}, where $\mathcal{O}(a^2)$ discretization effects were included in the 2-body quantization condition. The approach, however, only holds for discretization effects of certain types of fermionic actions, including e.g.\ Wilson, overlap, and domain-wall fermions.

Many \gls{lqcd} studies, however, use the staggered action in the fermion sector due to its high computational efficiency. Unfortunately, the quantization condition developed in Ref.~\cite{Hansen:2024cai} does not apply for staggered lattice artifacts, except (possibly) for very fine lattices, where these effects are small (see Ref.~\cite{Fu:2016itp,Fu:2024hhy} for studies where the Lüscher formalism was applied in the staggered case, neglecting discretization effects). In general, large staggered discretization effects hinder the extraction of scattering amplitudes from staggered \gls{lqcd} data at non-zero lattice spacing, the reason for which we briefly discuss below.

In the staggered theory, the Dirac operator is diagonalized exactly in spin space, allowing for a reduction of the fermion doubling from 16 to 4 copies. In the meson sector, the four remaining quark doublers (so-called tastes) lead to a pion multiplet (pion tastes) with unequal masses. These tastes cannot be interpreted as particle flavors. To obtain the correct number of flavors in the continuum limit, the fourth root trick is applied\footnote{The validity and robustness of this technique have been the subject of intense discussion in past decades. However, by now most of the issues around the fourth rooting trick have been addressed by lattice investigations, and the rooted staggered theory is believed to be under theoretical control. For a review on the discussion, see e.g.\ Ref.~\cite{Sharpe:2006re}.}. Altogether, the staggered \gls{lqcd} partition function reads
\begin{equation}
\mathcal{Z} = \int\mathcal{D}U\prod_f[\det(\slashed{D}+m_f)]^{1/4}\,.
\end{equation}
A notorious feature of this theory is that the rooted determinant cannot be associated with any known local action in terms of fermion degrees of freedom. As a result, the unitarity of the theory is broken\footnote{For comparison, we note that an analogous violation of unitarity also manifests in partially quenched theories where sea quark masses differ from the valence quark masses.}. Moreover, the existence of a pion multiplet suggests the possibility of more scattering channels than just that of the physical pion. However, these artifacts are of $\mathcal{O}(a^2)$ and thus vanish in the continuum limit.

Inspired by the developments in Ref.~\cite{Hansen:2024cai}, the goal of this work is to present the first steps towards resolving the difficulty around extracting scattering amplitudes from staggered \gls{lqcd} data at non-zero lattice spacing. In particular, we focus on extracting $\pi\pi$ scattering amplitudes below the $K\bar{K}$ threshold. To this end, we explore two complementary approaches. In the first approach, we use \gls{rschipt} to compute one-loop amplitudes for $\pi\pi$ scattering in the isospin-2 ($I=2$) case, where one does not expect unitarity-violating effects in the $s$-channel. Besides shedding light on the lattice artifacts coming from taste splitting and fourth rooting, these amplitudes can also point towards possible enhanced finite-volume effects in the quantization condition. In the second approach, we ultimately aim at generalizing the Lüscher formalism for $\pi\pi$ scattering to account for the dominant $\mathcal{O}(a^2)$ lattice artifacts associated with staggered fermions.

Given the vast amount of staggered lattice data available -- in particular, the Highly Improved Staggered Quarks (HISQ) ensembles by the FNAL/MILC collaboration, covering various volume sizes, pion masses, and lattice spacings -- we consider pursuing a formalism to extract scattering amplitudes from these data a timely endeavor.

\section{$\pi\pi$ scattering in Rooted Staggered Chiral Perturbation Theory}\label{sec:rschipt}

\subsection{S$\chi$PT at leading order}

In this subsection, we present a brief overview of \gls{schipt} to set the ground for our \gls{lo} and \gls{nlo} results for the $I=2$ ($\pi^+\pi^+\to\pi^+\pi^+$) scattering amplitudes. Following Ref.~\cite{Aubin:2003mg}, the Euclidean Lee-Sharpe Lagrangian of \gls{schipt} generalized to 3 flavors can be written as
\begin{equation}
\mathcal{L} = \frac{f^2}{8}\Tr[\partial_{\mu}\Sigma\partial_{\mu}\Sigma^{\dagger}] - \frac{\mu_0 f^2}{4}\Tr(M\Sigma + M\Sigma^{\dagger}) + \frac{2}{3}m_0^2(U_I + D_I + S_I)^2 + a^2\mathcal{U} + ...\,,
\label{eq:schipt_lagrangian}
\end{equation}
where the ellipsis indicates higher-order operator terms. $f$ and $\mu_0$ are \glspl{lec}, $m_0$ is a mass term due to the axial U(1) anomaly, $a$ is the lattice spacing, and $\mathcal{U}$ is the so-called taste-breaking potential. This potential is constructed from operators that violate taste symmetry, thus giving the pion tastes different masses. The matrices $\Sigma$ and $M$ are defined as
\begin{equation}
\Sigma = e^{i\Phi/f}\,,\hspace{0.5cm}\Phi = \mqty(U & \pi^+ & K^+ \\ \pi^- & D & K^0 \\ K^- & \bar{K}^0 & S)\,,\hspace{0.5cm}\text{and}\hspace{0.5cm}M = \mqty(m_u\mathds{1} & 0 & 0\\0 & m_d\mathds{1} & 0\\0 & 0 & m_s\mathds{1})\,,
\end{equation}
where $\mathds{1}$ is the $4\times4$ identity matrix, and each entry in the $\Phi$ matrix is a $4\times4$ matrix in taste space given by $\Phi_{ij} = \sum_{a=1}^{16}\Phi_{ij}^aT^a$, with $T_a\in\{\mathds{1},i\gamma_5\gamma_\mu,\gamma_\mu,\gamma_5,i\gamma_{\mu\nu}(\mu<\nu)\}$. As a result of taste splitting, the masses of each pion taste differ by lattice artifacts and, at first order, $m_{\pi,\xi}^2 = m_{\pi}^2 + a^2\Delta_\xi$, where $\xi\in\{I,A,V,P,T\}$ labels the taste, $m_{\pi}$ is the continuum pion mass, and $\Delta_\xi$ is the mass splitting, which can be computed using \gls{lqcd} (see e.g.\ Ref.~\cite{MILC:2004qnl}). The pion propagator in this theory is given by $D_\xi(k) = (k^2 + m_{\pi,\xi}^2)^{-1} - \delta_\xi(k)$, where the first term is the connected piece. $\delta_\xi(k)$ is a taste-dependent disconnected piece that gives the strength of the so-called hairpin vertex that appears for $\xi=A,V,I$ for flavor-neutral propagators. In our calculation, we consider $2+1$ flavors, where $u$ and $d$ are mass-degenerate. In this limit, the \gls{lo} amplitude for $I=2$ can be written schematically as
\begin{equation}
\M_{LO} \equiv \M_{\rm LO}(\pi_\xi^+\pi_\xi^+\to\pi_{\xi^{\prime}}^+\pi_{\xi^{\prime}}^+) = c_{\xi\xi^{\prime}}\M^{\rm cont}_{\rm LO}(m_{\pi}^2,s) + a^2F^{\xi\xi^{\prime}}_0\,,
\end{equation}
where $s\equiv (p_1+p_2)^2$ is the incoming momentum squared, $c_{\xi\xi'} \equiv \Tr[\gamma_\xi\gamma_\xi'\gamma_\xi\gamma_\xi']/4$ and $F_0^{\xi\xi^{\prime}}$ are taste-dependent coefficients that only depend on $f$ and on the \glspl{lec} in the taste-breaking potential $\mathcal{U}$ (c.f.\ Eq.~\eqref{eq:schipt_lagrangian}).  Notice that the effect of taste splitting becomes manifest already at tree level, driving the scattering amplitude away from the continuum result ($\M_{LO}^{\rm cont}$). However, in the continuum limit, we have $\lim_{a\to0}\M_{LO} = \pm\M_{LO}^{\rm cont}$. 

\subsection{Isospin-2 $\pi\pi$ scattering at next-to-leading order}

\begin{figure}[!ht]
    \centering
    \begin{subfigure}{0.28\textwidth}    \includegraphics[width=\linewidth]{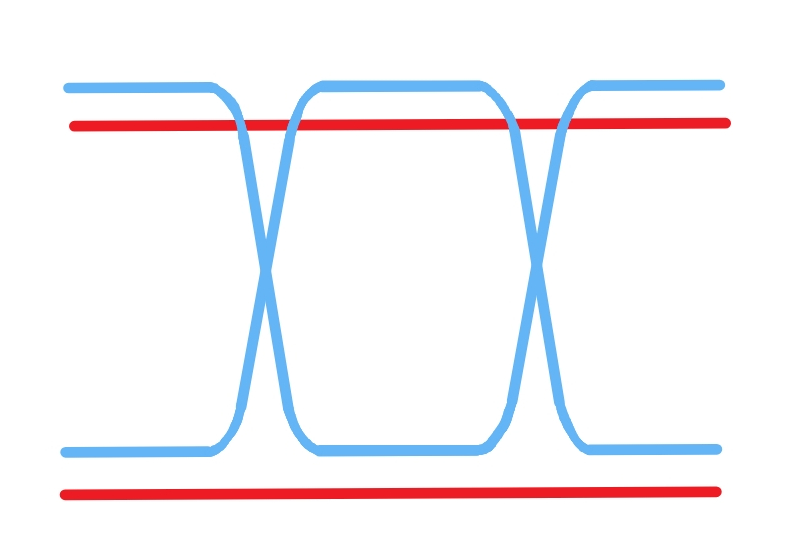}
    \caption{$s$-channel}
    \label{fig:diagrams_pipi_I2_a}
    \end{subfigure}
    \begin{subfigure}{0.28\textwidth}
    \includegraphics[width=\linewidth]{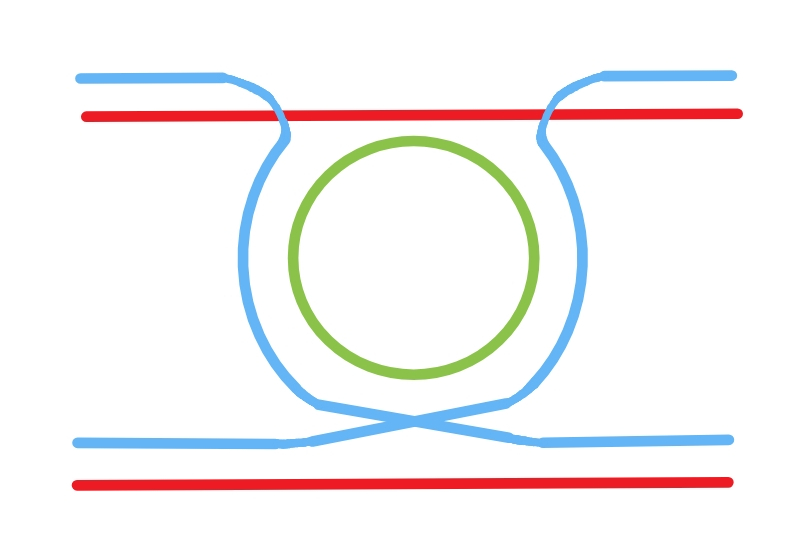}
    \caption{$t$-channel 1}
    \label{fig:diagrams_pipi_I2_b}
    \end{subfigure}
    \begin{subfigure}{0.28\textwidth}
    \includegraphics[width=\linewidth]{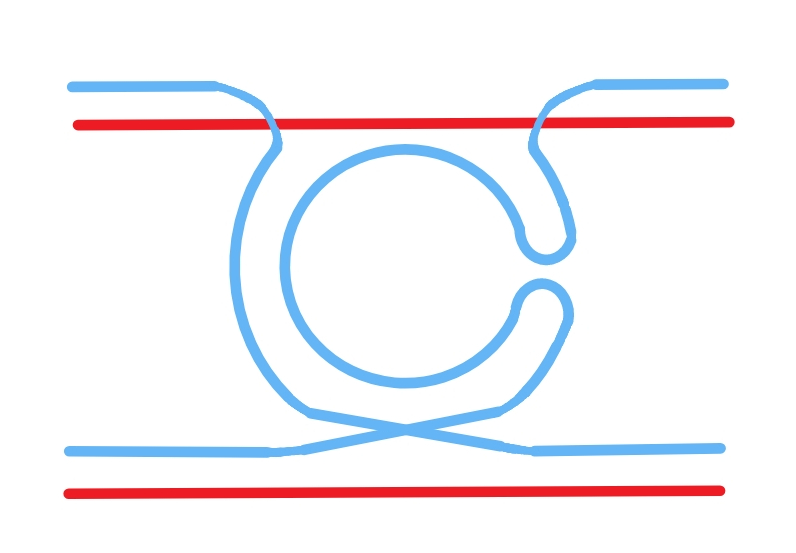}
    \caption{$t$-channel 2}
    \label{fig:diagrams_pipi_I2_c}
    \end{subfigure}
    \caption{(a) Example of $s$-channel diagram contributing to the $I=2$ $\pi\pi$ scattering. (b,c) Two examples of $t$-channel diagrams that are relevant for this scattering. The diagram (b) has an internal taste loop (inner green line) and requires a rooting factor (thus violating unitarity), whereas the diagram (c) involves disconnected propagators due to the hairpin vertex.}
    \label{fig:diagrams_pipi_I2}
\end{figure}

In Fig.~\ref{fig:diagrams_pipi_I2}, we show a few relevant one-loop diagrams in the $s$- and $t$-channels, whose amplitudes we schematically write down here in the isospin limit and for equal sea and valence quark masses.

For convenience, we use the finite-volume integrals; see e.g.\ Ref.~\cite{Bernard:2017scg} for their definition.
\begin{equation}
\mathcal{A}(m^2) \equiv -\int_V\frac{\dd^4 k}{(2\pi)^4}\frac{1}{k^2+m^2}\,, \hspace{0.5cm}{\cal B}(m_1^2,m_2^2,p^2) \equiv \int_V\frac{\dd^4 k}{(2\pi)^4}\frac{1}{(k^2+m_1^2)[(p-k)^2 + m_2^2]}\,.
\end{equation}
With these definitions, the $I=2$ $\pi\pi$ amplitude in the $s$-channel at \gls{nlo} can be expressed as
\begin{align}
\M_{\rm NLO}^{\rm s\,channel} = \frac{1}{16}\sum_{\Xi}&\left\{\qty[c_{\Xi \xi}c_{\Xi \xi^{\prime}}F_1(m_{\pi}^2,s) + a^2G^{\xi\xi^{\prime}\Xi}_1(m_{\pi}^2,s)]\mathcal{A}(m_{\pi,\Xi}^2)\right. \nonumber\\
+&\left.\qty[c_{\Xi \xi}c_{\Xi \xi^{\prime}}F_2(m_{\pi}^2,s) + a^2G^{\xi\xi^{\prime}\Xi}_2(m_{\pi}^2,s)]\mathcal{B}(m_{\pi,\Xi}^2,m_{\pi,\Xi}^2,s)\right\}
\end{align}
where $\Xi$ denotes the meson taste running in the loop. The functions $F_{1,2}$ and $G^{\xi\xi^{\prime}\Xi}_{1,2}$ are polynomial in their arguments and the $a^2$ functions $G_{1,2}^{\xi\xi'\Xi}$ also depend on the \glspl{lec} in $\mathcal{U}$. We note that this amplitude -- whose associated diagram is shown in Fig.~\ref{fig:diagrams_pipi_I2_a} -- includes $a^2$ corrections generated by both single-trace and double-trace operators in the taste-breaking potential. Analogously, the corresponding amplitudes for the $t$-channel diagram shown in Fig.~\ref{fig:diagrams_pipi_I2_b} is given by
\begin{align}
 \M_{\rm NLO}^{(b)} = \frac{1}{16}\sum_{\Xi}\frac{1}{4}\sum_{q=U,D,S}&\delta_{\xi\xi'}\left\lbrace \left[\widetilde{F}_1(m_\pi^2,s,t) + a^2 \widetilde{G}_1^{\xi\Xi}(m_\pi^2,s,t)\right] \mathcal{A}(m_{ql,\Xi}^2)\right.
  \nonumber\\ 
 &+\left.\left[\widetilde{F}_2(m_\pi^2,m_{ql}^2,s,t) + a^2 \widetilde{G}_2^{\xi\Xi}(m_\pi^2,m_{ql}^2,s,t)\right] \mathcal{B}(m_{ql,\Xi}^2,m_{ql,\Xi'}^2,t)\right\rbrace\nonumber\\
+\frac{1}{16}\frac{1}{4}\sum_{q=U,D,S}&(1-\delta_{\xi\xi'})\left\lbrace \left[c_{\xi\xi'}\widetilde{F}_1(m_\pi^2,s,t) + a^2 \widetilde{G}_3^{\xi\xi'}(m_\pi^2,s,t)\right]\mathcal{A}(m_{ql,\xi})\right.
  \nonumber\\ 
  &+\left.\left[c_{\xi\xi'}\widetilde{F}_2(m_\pi^2,m_{ql}^2,s,t)  + a^2 \widetilde{G}_4^{\xi\xi'}(m_\pi^2,m_{ql}^2,s,t)\right] \mathcal{B}(m_{ql,\xi}^2,m_{ql,\xi'}^2,t)\right.\nonumber\\
 &+ \left.\xi\leftrightarrow \xi'\right\rbrace\,,
\label{eq:t-channel_diagram_1}
\end{align}
where $t\equiv(p_1-p_3)^2$, with $p_1(p_3)$ the momentum of one of the incoming (outgoing) pions, and $m_{ql}=\mu_0(m_q+m_l)$ the LO mass with $m_l=m_u=m_d$. Notice that we made the rooting factor explicit in the expression above. Finally, for the diagram in Fig.~\ref{fig:diagrams_pipi_I2_c} 
\begin{align}
  \M_{\rm NLO}^{(c)} = \frac{1}{16}
  \sum_{\Xi=A,V,I} \delta_{\xi\xi'}\sum_{m_P=m_\pi,m_\eta,m_{\eta'}}
\left\lbrace \left[\hat{F}_{1,P}^\Xi(m_\pi^2,m_\eta^2,t) + a^2 \hat{G}_{1,P}^{\xi\Xi}(m_\pi^2,m_\eta^2,t)\right] \mathcal{A}(m_{P\Xi}^2)\right.\nonumber\\
\left.+\left[\hat{F}_{2,P}^\Xi(m_\pi^2,m_\eta^2,t) + a^2 \hat{G}_{2,P}^{\xi\Xi}(m_\pi^2,m_\eta^2,t)\right] \mathcal{B}(m^2_{\pi\Xi},m^2_{P\Xi},t)\right\rbrace\nonumber\\
+  \frac{1}{16}
  \sum_{\Xi=A,V,I} (1-\delta_{\xi\xi'})\left\lbrack \delta_{\Xi\xi'}\sum_{m_P=m_\pi,m_\eta,m_{\eta'}}
\left\lbrace \left[\hat{F}_3^{\xi'}(m_\pi^2,m_\eta^2,t) + a^2 \hat{G}_3^{\xi\xi'}(m_\pi^2,m_\eta^2,t)\right] \mathcal{A}(m_{P\xi'}^2)\right.\right.\nonumber\\
\left.+\left[\hat{F}_4^{\xi'}(m_\pi^2,m_\eta^2,t) + a^2 \hat{G}_4^{\xi\xi'}(m_\pi^2,m_\eta^2,t)\right] \mathcal{B}(m^2_{\pi\xi},m^2_{P\xi'},t)\right\rbrace\nonumber\\
\left.\left.+\left[\hat{F}_5^{\xi´}(m_\pi^2,m_\eta^2,t) + a^2 \hat{G}_5^{\xi\xi'}(m_\pi^2,m_\eta^2,t)\right] \mathcal{A}(m_{\pi\xi}^2)\right\rbrace + \xi\leftrightarrow \xi'\right\rbrack\,,
  \label{eq:t-channel_diagram_2}
\end{align}
where the sum over $m_P$ excludes $m_\eta'$ for $\Xi=I$. The functions $\widetilde{G}_j^{\alpha\beta}$ and $\hat{G}_j^{\alpha\beta}$ in Eqs.~(\ref{eq:t-channel_diagram_1}) and (\ref{eq:t-channel_diagram_2}), respectively, also depend on \glspl{lec} in  the taste-breaking potential $\mathcal{U}$. Notice that the sum over $\Xi$ in Eq.~\eqref{eq:t-channel_diagram_1} runs over all tastes, whereas in Eq.~\eqref{eq:t-channel_diagram_2} it only runs over tastes with a hairpin vertex.

It is worth mentioning that since the diagram in Fig.~\ref{fig:diagrams_pipi_I2_c} contains a disconnected propagator, all $\hat F_j^\beta$ and $\hat G_j^{\alpha\beta}$ functions in Eq.~\eqref{eq:t-channel_diagram_2} are proportional to the corresponding hairpin parameter of a given taste $\beta$, that is $\mathcal{O}(a^2)$ for $\beta=A,V$. Hence, in fact, the G functions give $\mathcal{O}(a^4)$ contributions for $\Xi,\xi'=A,V$. We will provide these functions alongside other details of the calculation and similar results for the $I=0,1$ channels in a future publication. We stress that the $t$-channel of this scattering process contains various diagram topologies, which are similar to the ones shown here, as opposed to the $s$-channel, which contains only one topology.

Since these amplitudes are related to each other via analytic continuation, we expect that for other isospin cases, their corresponding $s$-channels contain multiple diagram topologies, similarly to the $t$-channel of the isospin-2 case. Although we do not show amplitudes explicitly in these cases, this observation will be relevant in the next section, when we discuss a possible route to generalize the Lüscher formalism to include staggered lattice artifacts.

\section{Towards a Lüscher formalism with staggered lattice artifacts}

In this section, we discuss our second approach, where we present the first steps towards modifying the Lüscher formalism for staggered \gls{lqcd}. Specifically, we argue that fourth-rooting and taste-splitting effects might be incorporated by means of (at least) three modifications of the original formalism, which we dub: (1) introducing multiple $s$-channel topologies, (2) rescaling $s$-channel diagrams, and (3) considering a multichannel system. The third modification is, in fact, already known in the flavor case~\cite{Briceno:2012yi}, which we adapt here to the multi-taste case. To illustrate our approach, we closely follow the derivation of the quantization condition via the so-called skeleton expansion as in Ref.~\cite{Hansen:2024cai} and introduce our modifications in the intermediate steps.  

\subsection{\boldmath Multiple $s$-channel topologies}

It is well known that $s$-channel scattering is the most relevant one for the Lüscher formalism due to the possibility of particles going on-shell~\cite{Luscher:1986pf}. In Sec.~\ref{sec:rschipt}, we noted using one-loop \gls{rschipt} that a single $s$-channel diagram topology contributes to $\pi\pi$ scattering in the $I=2$ case, and argued that multiple topologies contribute to the $I=0,1$ cases. Therefore, to obtain a formalism for different isospin cases, it is necessary to account for the possibility of a generic number of $s$-channel diagram topologies in the skeleton expansion. This corresponds to our first modification of the formalism.

\begin{figure}[!ht]
    \centering
    \begin{subfigure}{0.21\textwidth}
    \caption*{$\pi^+\pi^+ \longrightarrow\pi^+\pi^+$}
    \includegraphics[width=\linewidth]{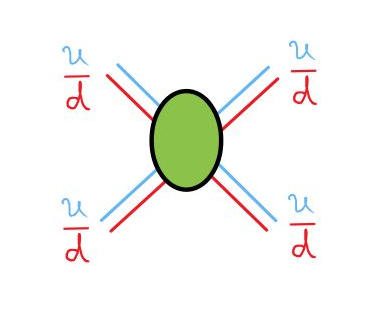}       
    \end{subfigure}
    \begin{subfigure}
    {0.21\textwidth}
    \caption*{$\pi^+\pi^- \longrightarrow\pi^+\pi^-$}
    \includegraphics[width=\linewidth]{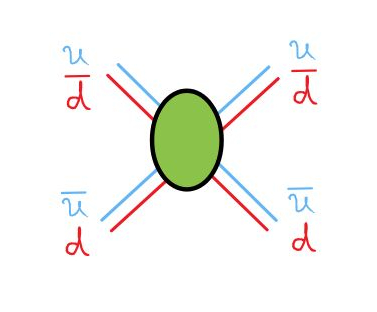}       
    \end{subfigure}
    \begin{subfigure}
    {0.21\textwidth}
    \caption*{$\pi^0\pi^0 \longrightarrow\pi^0\pi^0$}
    \includegraphics[width=\linewidth]{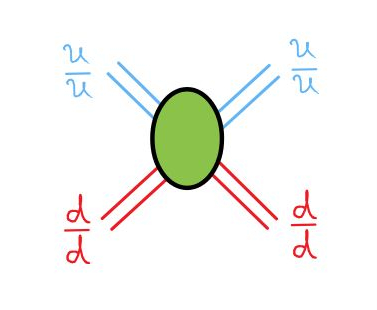}       
    \end{subfigure}
    \caption{From left to right, Bethe-Salpeter kernels for the processes $\pi^+\pi^+\to\pi^+\pi^+$ ($I=2$), $\pi^+\pi^-\to\pi^+\pi^-$, and $\pi^0\pi^0\to\pi^0\pi^0$. As usual, the green blobs represent a sum of all diagrams that are not in the $s$-channel.}
    \label{fig:bs_kernels}
\end{figure}

To begin with, we illustrate the so-called \gls{bs} kernels for the scattering of pions entering the various isospin channels in Fig.~\ref{fig:bs_kernels}.
The scattering amplitude with $N$ $s$-channel diagram topologies is defined via the formal relation
\begin{equation}
\M = B + \sum_{n=1}^NB\cdot\I_n\cdot\M\,.
\label{eq:formal_scattering_amplitude}
\end{equation}
where $B$ is a given \gls{bs} kernel, which is a function of the incoming and outgoing momenta. The integral operator $\I_n$ is defined analogously to Ref.~\cite{Hansen:2024cai}
\begin{equation}
L\cdot \I_n \cdot R \equiv \int\frac{\dd^3\vec{k}}{(2\pi)^3}\int\frac{\dd k_4}{2\pi}L(k)D_n(k)D_n(P-k)R(k)\,,
\label{eq:loop_integral_operator}
\end{equation}
where $D_n$ is the bare propagator of the particle running in the diagram $n$, and $L$ and $R$ are arbitrary functions of the momentum $k=(\vec{k},k_4)$. $P$ is the incoming momentum of the two pions. In the \gls{cm} frame, for instance, $P=(\vec{0},E)$, where $E$ is the 2-pion energy. Inserting Eq.~\eqref{eq:formal_scattering_amplitude} into itself multiple times, we get
\begin{align}
\M = B + \sum_{n=1}^{N}B\cdot \I_n\cdot B + \sum_{n=1}^{N}\sum_{n^{\prime}=1}^{N}B\cdot \I_n\cdot B\cdot \I_{n^{\prime}}\cdot B + ...\,,
\label{eq:M_geometric_series}
\end{align}
Before we derive the quantization condition, it is convenient to introduce the second modification of the formalism.

\subsection{\boldmath Rescaling $s$-channel diagrams}

\begin{figure}[!ht]
    \centering    \includegraphics[width=\textwidth]{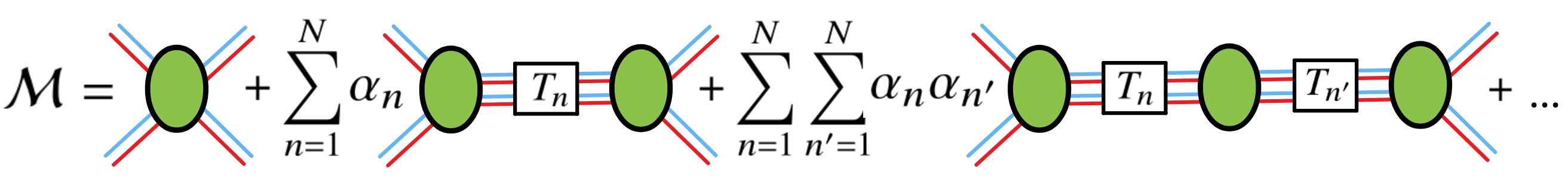}
    \caption{Modified skeleton expansion to include $N$ $s$-channel topologies $T_n$ rescaled by an arbitrary factor $\alpha_n$. This factor is set to $1/4$ if the diagram contains an internal taste loop, or to $1$ otherwise.}
    \label{fig:M_skeleton_expansion}
\end{figure}

Another aspect of \gls{rschipt} highlighted in Sec.~\ref{sec:rschipt} is the need to multiply diagrams with internal taste loops by $1/4$. This ad hoc procedure implements the fourth rooting in a perturbative fashion. Therefore, to introduce fourth-rooting effects into the formalism, we modify the skeleton expansion to allow for a rescaling of $s$-channel diagrams by a factor $\alpha_n$, which we will later match to the rooting prescription. After replacing $\I_n\to\alpha_n\I_n$ in Eq.~\eqref{eq:M_geometric_series}, the result reads
\begin{align}
\M = B + \sum_{n=1}^{N}\alpha_nB\cdot \I_n\cdot B + \sum_{n=1}^{N}\sum_{n^{\prime}=1}^{N}\alpha_n\alpha_{n^{\prime}}B\cdot \I_n\cdot B\cdot \I_{n^{\prime}}\cdot B + ...\,.
\label{eq:scattering_amp_multi_loop}
\end{align}
We note that the rescaling procedure can be seen as a rescaling of each propagator in Eq.~\eqref{eq:loop_integral_operator} by a factor $\sqrt{\alpha_n}$. In Fig.~\ref{fig:M_skeleton_expansion}, we depict the modified skeleton expansion for multiple $s$-channel diagrams, including the rescaling by an arbitrary factor.

To derive the quantization condition, we perform an analogous expansion for the finite-volume correlator $G_L$, with the difference that the integral operators $\I_n$ are replaced by sum operators $\S_n$, which are defined by replacing $\displaystyle(2\pi)^{-3}\int\dd k^3\longrightarrow L^{-3}\sum_{\vec{k}}$ in Eq.~\eqref{eq:loop_integral_operator}. The result is
\begin{align}
G_L = B + \sum_{n=1}^{N}\alpha_nB\cdot \S_n\cdot B + \sum_{n=1}^{N}\sum_{n^{\prime}=1}^{N}\alpha_n\alpha_{n^{\prime}}B\cdot \S_n\cdot B\cdot \S_{n^{\prime}}\cdot B + ...\,.
\label{eq:finite_volume_correlator_expansion}
\end{align}
Finally, we define $F_n\equiv\S_n-\I_n$, where (up to exponentially suppressed terms in the volume) $F_n$ contains only on-shell propagators\footnote{This is known to be true for connected propagators. In the staggered case, where some tastes have disconnected (hairpin) propagators, this must be further investigated.}. Expressing $\S_n$ in terms of $F_n$ on every instance in Eq.~\eqref{eq:finite_volume_correlator_expansion}, we can show that the dependence on the \gls{bs} kernel can be eliminated. The final result reads
\begin{equation}
G_L = \frac{1}{\M^{-1} - \sum_{n=1}^{N}\alpha_n(\S_n-\I_n)}\,.
\label{eq:finite_volume_correlator_no_bs_kernel}
\end{equation}
Thus, the poles of $G_L$ appear for energies that satisfy the following condition
\begin{equation}
\det[\M^{-1}(E) - F_{\rm tot}(E,L)] = 0\,,
\label{eq:quantization_condition}
\end{equation}
where $F_{\rm tot} \equiv \sum_{n=1}^N\alpha_s(\S_n-\I_n)$ is such that
\begin{equation}
L\cdot F_{\rm tot} \cdot R = \sum_{n=1}^N\alpha_n\qty[\frac{1}{L^3}\sum_{\vec{k}}-\int\frac{\dd^3\vec{k}}{(2\pi)^3}]\int\frac{\dd k_4}{2\pi}L(k)D_n(k)D_n(P-k)R(k)\,.
\label{eq:finite_volume_object}
\end{equation}
To make contact with the rooting prescription of \gls{rschipt} (c.f.\ sec.~\ref{sec:rschipt}), we can set $\alpha_n=1/4$ if the diagram $n$ contains internal taste loops, or $\alpha_n=1$ otherwise. In such a way, the rooting procedure can be included in the quantization condition at all orders in perturbation theory. Moreover, notice that introducing unequal factors $\alpha_n$ for each diagram breaks unitarity in the desired fashion. Likewise, one could also use Eq.~\eqref{eq:finite_volume_object} to describe the unrooted staggered case by setting all $\alpha_n$ equal to unity. It is important to stress that the \gls{bs} kernels shown in Fig.~\ref{fig:M_skeleton_expansion} contain all diagrams that are not in the $s$-channel, including e.g.\ $t$-channel and $u$-channel diagrams, thus incorporating unitarity violation in all isospin cases. This must be taken into account when deriving matrix elements for the operator $F_{\rm tot}$ from Eq.~\eqref{eq:finite_volume_object}.

\subsection{Multichannel system}
Since the staggered theory has multiple pions in its spectrum, it is reasonable to expect that the scattering of two pions with fixed taste couples to that of pions with different tastes, as long as the total taste quantum number is conserved. Hence, we consider a multichannel system with all possible scattering channels (for details on the multichannel formalism, see Ref.~\cite{Briceno:2012yi} and references therein).

For convenience, let us assume without loss of generality that the two pions in the initial state have the same taste $\xi$, thus forming a taste singlet. By total taste conservation, the two pions in the final state must also have the same taste, which we denote by $\xi^{\prime}$, such that the scattering process is given by $\pi_{\xi}\pi_{\xi}\to\pi_{\xi^{\prime}}\pi_{\xi^{\prime}}$. Let $\M_{\xi\to \xi^{\prime}}$ denote the scattering amplitude for this process. The full scattering matrix, considering all allowed taste channels, is then given by
\begin{equation}
\M = \mqty(\M_{I \to I} & \M_{I \to V} & \M_{I \to A} & \M_{I \to P} & \M_{I \to T} \\ \M_{V \to I} & \M_{V \to V} & \M_{V \to A} & \M_{V \to P} & \M_{V \to T} \\ \M_{A \to I} & \M_{A \to V} & \M_{A \to A} & \M_{A \to P} & \M_{A \to T} \\ \M_{P \to I} & \M_{P \to V} & \M_{P \to A} & \M_{P \to P} & \M_{P \to T} \\ \M_{T \to I} & \M_{T \to V} & \M_{T \to A} & \M_{T \to P} & \M_{T \to T})\,.
\label{eq:scattering_matrix_multichannel}
\end{equation}
Notoriously, the number of unknown variables increased considerably compared to the single-channel case. A common approach to solving for the different scattering amplitudes in this situation is to model\footnote{For a recently proposed strategy, where the scattering amplitudes are extracted in a model-independent fashion, see Ref.~\cite{Salg:2025now}} them using physically motivated \textit{Ansätze}~\cite{Guo:2012hv}. To constrain the model parameters, one requires a sufficiently large number of energies for various momenta and across multiple lattice volumes. In Ref.~\cite{Lahert:2024vvu}, the 2-pion staggered energies were calculated in the $I=1$ case at zero total momentum on a coarse ensemble. These calculations will be extended for other isospin channels, non-zero momenta, and finer ensembles.

\section{Conclusions and outlook}
Due to their high computational efficiency, rooted staggered fermions play a fundamental role in modern \gls{lqcd} calculations -- with applications ranging from QCD thermodynamics~\cite{Borsanyi:2025ttb} to the anomalous magnetic moment of the muon~\cite{Aliberti:2025beg}. Nevertheless, the lattice artifacts of the rooted staggered theory impose several theoretical roadblocks to the extraction of scattering amplitudes via the Lüscher formalism. Here, we presented the first steps to tackle these roadblocks employing two complementary approaches.

In the first approach, we computed for the first time $\pi\pi$ scattering amplitudes in the $I=2$ case using \gls{rschipt} at one loop. We highlighted several aspects of our results that will provide guidance when implementing staggered lattice artifacts into the Lüscher formalism, for instance, the existence of multiple diagram topologies and the analogy between different tastes and the multichannel formalism. In the second approach, we proposed a way to include these different diagram topologies, fourth-rooting effects, and multiple tastes into the quantization condition.

In the future, we will use our one-loop \gls{rschipt} results to explicitly check the validity of the quantization condition for all isospin cases (for a similar check in partially quenched \acrshort{chipt}, see Ref.~\cite{Draper:2021wga}). We also aim to calculate energy levels from these amplitudes~\cite{Bernard:1995ez} to investigate their volume dependence and how they are modified by the staggered lattice artifacts. Another feasible approach, which we plan to explore in the future, is to take the continuum limit of the energy levels at fixed volume first and then apply the standard version of the Lüscher formalism. In this limit, the $\mathcal{O}(a^2)$ staggered lattice artifacts vanish, and the theory is unitary. The challenge in this case is to keep the volume constant while taking the continuum limit, since the lattice volumes are typically not the same across different lattice ensembles.

\acknowledgments

The authors thank Maxwell T. Hansen, André Walker-Loud, Fernando Romero-López, Stephen Sharpe, and Maarten Golterman for productive and insightful discussions that guided this work substantially. This work was partially supported by 
MICIU/AEI/10.13039/501100011033 and FEDER (EU) under Grant PID2022-140440NB-C21 and by Consejería de Universidad, Investigación y Innovación and Gobierno de España and EU -- NextGenerationEU, under Grant AST22 8.4; 
 The U.S. Department of Energy, Office of Science, under Award No. DE-SC0015655 (A.X.K., M.D.) and the Funding Opportunity Announcement Scientific Discovery through Advanced Computing: High Energy Physics, LAB 22-2580 (S.L.); 
 and by the National Science Foundation under Grants Nos. PHY20-13064 and PHY23-10571 (S.L.).

\bibliographystyle{JHEP}
\bibliography{bibliography.bib}

\end{document}